\documentclass[9pt,twocolumn,twoside]{osajnl}
\journal{ol} % Choose journal (ao, aop, josaa, josab, ol, pr)

\setboolean{shortarticle}{true}
\title{Anti-parity-time topologically undefined state}

\author[1]{Haohao Wang}
\author[1,5]{Kaiwen Ji}
\author[1]{Yuandan Wang}
\author[1]{Zhenjuan Liu}
\author[2]{Yuanmei Gao}
\author[3]{Yanlong Shen}
\author[4]{Shi Bai}
\author[4]{Koji Sugioka}
\author[1*]{Xinyuan Qi}

\affil[1]{School of Physics, Northwest University, 710127, Xi'an, China}
\affil[2]{College of Physics and Electronics, Shandong Normal University, Jinan, 250014, China}
\affil[3]{State Key Laboratory of Laser Interaction with Matter, Northwest Institute of Nuclear Technology, Xi’an Shaanxi 710024, China}
\affil[4]{RIKEN Center for Advanced Photonics, Hirosawa 2-1, Wako, Saitama 351-0198, Japan}
\affil[5]{Centre for Nanosciences and Nanotechnologies, Université Paris-Saclay, Palaiseau, France}

\affil[*]{Corresponding author: qixycn@nwu.edu.cn}
\begin{abstract}
We constructed an anti-parity-time-symmetric photonic lattice by using perturbations. The results show the topological state will appear when the waveguide coupling constants $\kappa_1<\kappa_2$; Interestingly, a state with undefined winding numbers occurs when $\kappa_1=\kappa_2$, in which the light distributes only in the wide waveguides with equal magnitude distribution. Further studies show that the edge state will be strengthened by introducing defect for the topologically non-trivial case, while it will not affect the equal intensity transmission for the topologically undefined state. Our work provides a new way to realize the topological state and equally divided light transmission and might be applicable in optical circuits and optical interconnect.
\end{abstract}

\setboolean{displaycopyright}{true}

\begin{document}
	
	\maketitle

	\label{sec:examples}
	
	\section{Introduction}
	Topological photonics, as one of the most remarkable topics, is inspirated from the discovery of the quantum Hall effects and the topological insulators in the condensed matter~\cite{01Markus2007,02Petrifmmode2020}. In recent years, a variety of discrete photonic systems have been proved to realize the topological phase~\cite{03ozawa_topological_2019}, such as the Floquet topological insulators with spatial modulation based on the array of laser-written optical waveguides~\cite{04mukherjee_experimental_2017} and a network of surface plasmon  rings~\cite{05gao_probing_2016}, topological photonics in meta-waveguides~\cite{06cheng_robust_2016} and photonic topological insulators~\cite{07wu_scheme_2015}. Based on these studies, a large number of phenomena have emerged, such as the topologically protected edge states and corner states in photonic crystals\cite{08zhang2020}, slow light~\cite{09yu2021}, nonlinear optical isolation~\cite{10Zhou2017}. Clearly, a common feature of these studies is that the systems must be Hermitian, which means the topological invariant of a system with open boundary condition could be judged by the bulk-edge correspondence~\cite{11Chern1993Hatsugai}.
	
	In the non-Hermitian systems where there exists energy exchange with the ambient environment, the fractional winding number could be observed, and the usual bulk-edge correspondence is invalid~\cite{12Anomalous2016Lee}. The global Berry phase~\cite{13liangtopological2013}, the non-Bloch winding number~\cite{14Edge2018Yao,15Bulk2019Jin}, and the Kronecker index~\cite{16quandt_winding_2021} could be employed to calculate the winding numbers. These findings promote the development of topology research in the non-Hermitian system~\cite{03ozawa_topological_2019,17Wang2020,18Yao2018, 19Borgnia2020,20Sone2020,21Anomalous2020Gao}. The propagation of light in these systems will be affected by external conditions, leading to many novel phenomena, such as lasing generation in topological edge states of a polariton micropillar array~\cite{03ozawa_topological_2019}, fermi nodal disk in magnetic plasma\cite{17Wang2020}, skin effects~\cite{18Yao2018, 19Borgnia2020},  boundary modes~\cite{19Borgnia2020}, exceptional non-Hermitian topological edge mode\cite{20Sone2020}, and anomalous topological edge state~\cite{21Anomalous2020Gao} and so on.
	
	However, there exists a particular state with an undefined topological invariant in all of these studies. For the two-band model, once the closed parametric curve of the coefficient vector $\rm{\boldsymbol{d}}(k)$ of bulk momentum-space Hamiltonian $\boldsymbol{\hat{H}(k)}$ is tangent to the origin, the winding number is undefined~\cite{22Asboth2016}, which means that such a topologically undefined state is a critical state between the topologically non-trivial and topologically trivial ones. Considering the signs around the Dirac point in two-dimensional systems are opposite while the dispersion surface around the Weyl points is changed dramatically~\cite{23Lu2014,03ozawa_topological_2019}, most of the current studies are focused on the linear regions, and therefore the studies of light dynamics at these points are still remarkably absent.
	 
	In this Letter, we propose a one-dimensional anti-parity-time-symmetric (anti-PT) photonic lattice with perturbations. By eliminating the amplitudes of the perturbation sites adiabatically, we obtained an effective Hamiltonian with anti-PT symmetry. The system is proved theoretically to be topologically non-trivial when the coupling constants $\kappa_1<\kappa_2$; while there exists a topologically undefined state when $\kappa_1=\kappa_2$. It is also found that the introduction of the defect will strengthen the edge state for the topologically non-trivial case while it will not affect the equal intensity transmission for the topologically undefined state. Our study enriches the physics of topological photonics and paves a new way to achieve the topological edge state and undefined state, which might have potential applications in the fields of optical circuits and optical communication. 
	
	\section{Model and Analysis}
	
\begin{figure}[t]
		\centering
		\fbox{\includegraphics[width=\linewidth]{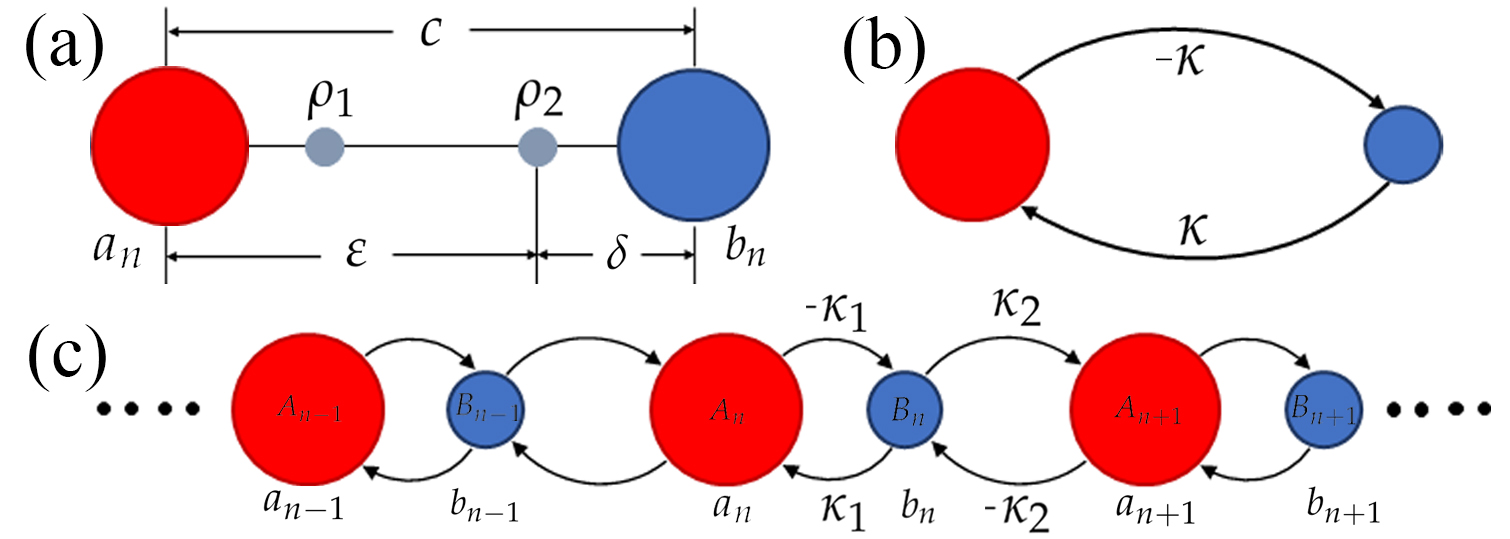}}
		\caption{Schematic of photonic lattices with perturbations. (a) Perturbations used to realize anti-PT symmetry. (b) Effective model. (c) Periodic anti-PT structure. $\rho_n$, the perturbation sites (the gray sites); $A_n$(the red sites) and $B_n$(the blue sites) are wide and narrow waveguides, respectively. $\kappa_{1, 2}>$ 0 are the coupling constants.}
		\label{fig.1}
	\end{figure}

	We construct a one-dimensional photonic lattice with perturbations, as shown in Fig.~\ref{fig.1}(a). $\rho_n$, $a_n$, and $b_n$ are the light amplitudes of the perturbation site, ordinary waveguides $A_n$ and $B_n$, respectively. Assuming that the refractive index of the perturbation site is much smaller than those on sites $A_n$ and $B_n$, $n_\rho \ll n_{a, b}$, therefore the modal amplitudes in the ordinary waveguides are more substantial than in the perturbations~\cite{24Ji2018}. Accordingly, the ordinary waveguide will not be affected by the next neighbor perturbation. Under the above conditions, the Hamiltonian of the periodic system, $H'$, with perturbations is described as
		
	\begin{gather}
		\label{eq:1}
		\boldsymbol{H'}=\begin{pmatrix}
			\beta_a & c & \delta & 0 \\
			c & \beta_b & 0 & \delta \\
			\delta & \varepsilon & \beta_{\rho,1} & 0\\
			\varepsilon & \delta & 0 & \beta_{\rho,2} \\
		\end{pmatrix}, 
	\end{gather}
	where $\delta$ and $\varepsilon$ are the nearest and next-nearest coupling constants between the perturbations and the ordinary waveguides. $c$ is the coupling constant between ordinary waveguides. $\beta$ is the propagation constant. One can describe the evolution of this system by the Schr\"{o}dinger-like equation~\cite{25Ji2020} $i \frac{d\boldsymbol{\psi'_n}}{dz}=\boldsymbol{H}' \boldsymbol{\psi}'_n$ and the wave functions of $\boldsymbol{\psi}'_n=[a_n, b_n, \rho_1, \rho_2]^{\boldsymbol{T}}$. 
	
	\begin{figure}[h]
	\centering
	\fbox{\includegraphics[width=\linewidth]{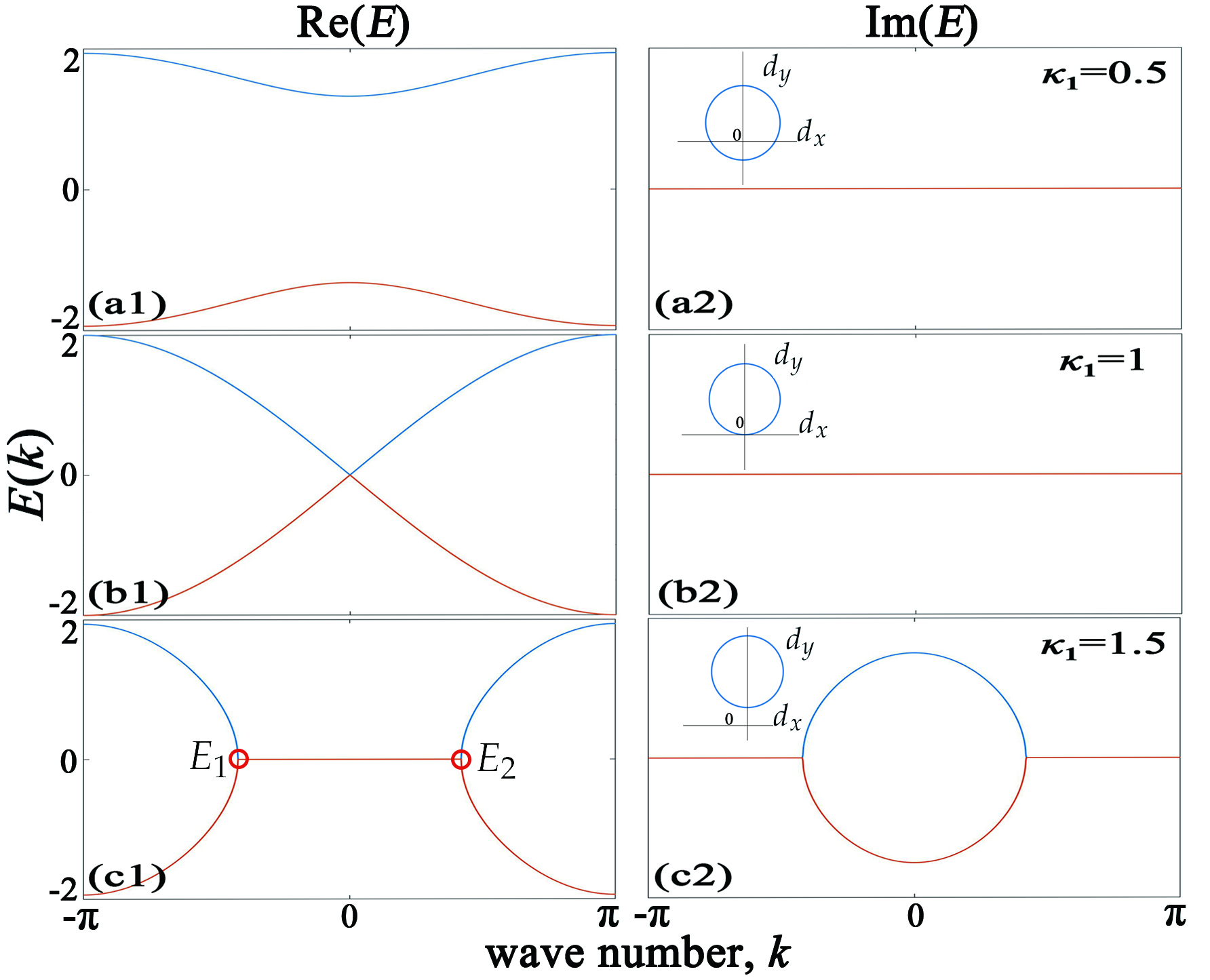}}
	\caption{Bulk band structure with different $\kappa_1$. }
	\label{fig.2}
    \end{figure}
	
	To simplify the equations, one can eliminate the amplitudes of perturbation sites $\rho_n$ adiabatically ~\cite{26PhysRevB.95.014201}, 
	\begin{gather}
		\label{eq:2}
		\rho_1\approx\frac{-a\delta-b\varepsilon}{\beta_{\rho,1}},
		\rho_2\approx\frac{-b\delta-a\varepsilon}{\beta_{\rho,2}}.
	\end{gather}
	
	After setting the propagation constants of the perturbation sites as $\beta_{\rho,1}=\delta\varepsilon/(c-\kappa), \beta_{\rho,2}=\delta\varepsilon/(c+\kappa)$ and those for the ordinary waveguides as $\beta_a=(c\delta+\Delta \varepsilon-\delta \kappa)/\varepsilon$ and $\beta_b=(c\delta-\Delta \varepsilon+\delta \kappa)/\varepsilon$, one can obtain the effective Hamiltonian
	\begin{equation}
		\label{eq:3}
		\boldsymbol{H}=
		\begin{pmatrix}
			\Delta & \kappa\\
			-\kappa & -\Delta
		\end{pmatrix}, 
	\end{equation}
	where $\Delta$ and $\kappa$ are the corresponding effective propagation constant and coupling constant, respectively. The simplified system is shown in Fig. \ref{fig.1}(b). 
	The effective Hamiltonian satisfies $\boldsymbol{PTH+HPT=0}$, where $\boldsymbol{P}=\sigma_x$ and $\boldsymbol{T}=IK$ are the parity and time operators, respectively. $\sigma_{x,y,z}$ are the Pauli matrices, $I$ is the identity operator, $K$ is the elementwise complex conjugation in the effective Hamiltonian. Apparently, the effective Hamiltonian of such a system satisfies anti-PT symmetry but is not chirally symmetric,  $\boldsymbol{\Gamma H\Gamma^{-1} \neq -H}$ , where $\boldsymbol{\Gamma} = \sigma_{z}$. 
	
	As shown in the aforementioned analyzation, we have realized an anti-PT symmetric optical coupler. Based on this, we employ such a configuration to construct a topological Su-Schrieffer-Heeger (SSH) model~\cite{03ozawa_topological_2019} [see Fig.~\ref{fig.1}(c)]. The Bloch Hamiltonian of such a periodic system is
	\begin{gather}
		\label{eq:4}
		\boldsymbol{H}=-id_x \sigma_x+id_y\sigma_y+\Delta\sigma_z,\\
		\label{eq:5}
		d_x=\kappa_2\sin k,\\
		\label{eq:6}
		d_y=\kappa_1+\kappa_2\cos k,
	\end{gather}
	where $\kappa_1$ and $\kappa_2$ are the coupling constants inside and outside one cell. One can learn that the Hamiltonian satisfies $\{\boldsymbol{H}, \tau\}=0$, where $\tau=\sigma_xK$,  in other words, the system has non-Hermitian particle-hole symmetry, which guarantees that the eigenvalues always appear in pair $\pm E(k)$. The corresponding eigenvalues of the bulk Hamiltonian read
	\begin{gather}
		\label{eq:7}
		E(k)=\pm\sqrt{\Delta^2-d_x^2-d_y^2}.
	\end{gather}
	
	The eigenvalues of the bulk Hamiltonian are demonstrated in Fig. \ref{fig.2}, one can notice that the two bands are closing to each other with the increasing $\kappa_1$, and a Dirac point is formed at $k=0$ when $\kappa_1=1$[see Fig.~\ref{fig.2}(b1)]. It is also shown in the inset in Fig.~\ref{fig.2}(b2) that the circle is tangent to the origin in the inset, this indicates that the winding number of the system is undefined. Further increasing the magnitude of the inner coupling constant $\kappa_1$, a flatband starts to emerge in the center of the Brillouin zone as a result of the protection of non-Hermitian particle-hole symmetry[see Fig.~\ref{fig.2}(c1)]. Such a flatband covers a region of $-\sqrt{d_x^2+d_y^2}<\Delta<\sqrt{d_x^2+d_y^2}$ in which $E_1$ and $E_2$ are also known as the exceptional points~\cite{28Xia2021}. We further calculated the winding number $W$ of such a system~\cite{25Ji2020}. It is shown $W =1$ corresponding to a topological non-trivial state when $\kappa_1<\kappa_2$, and $W=\frac{1}{2}$ corresponding to an undefined state when $\kappa_1=\kappa_2$.
		
	\begin{figure}%[h]
		\centering
		\fbox{\includegraphics[width=\linewidth]{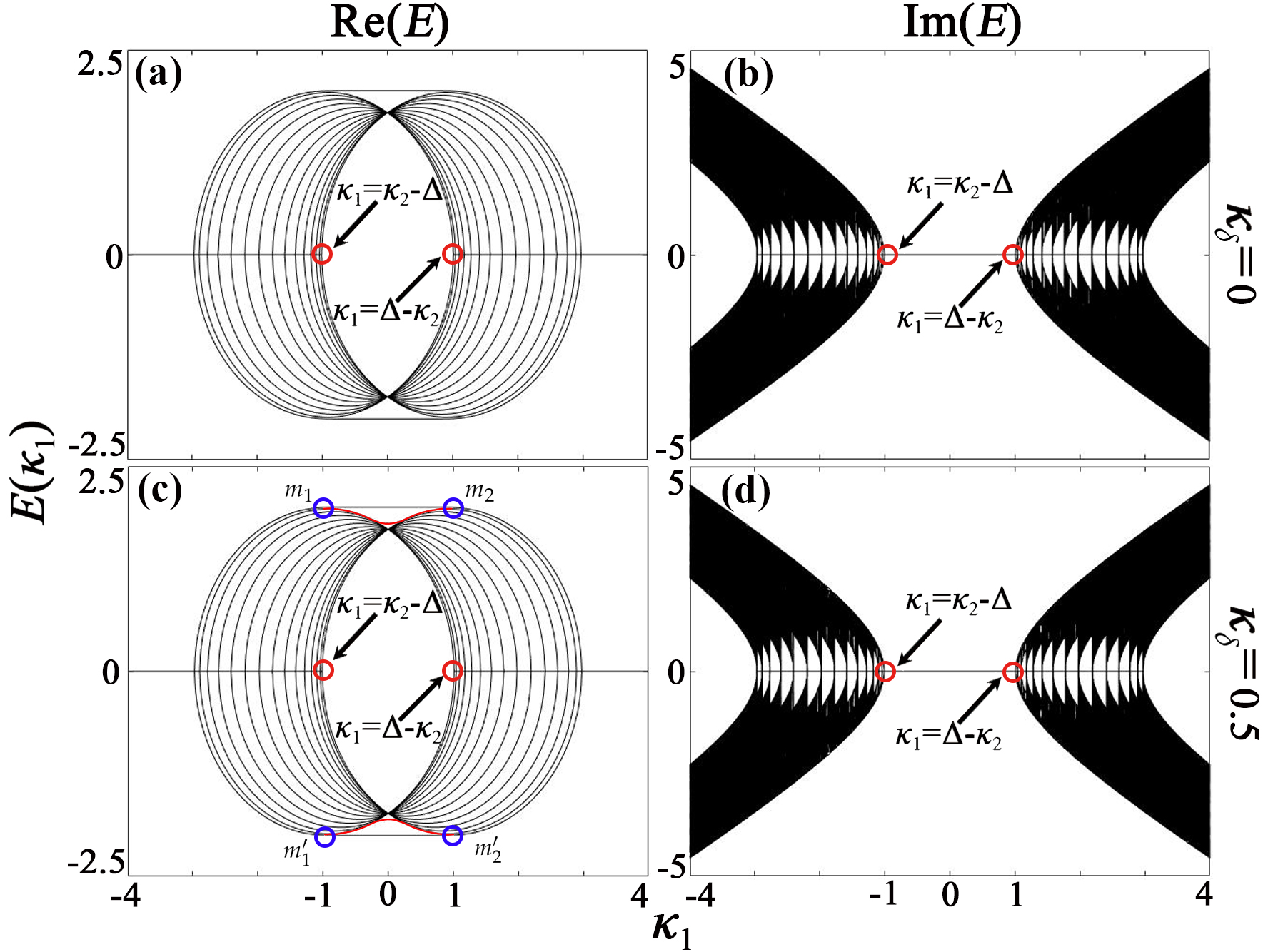}}
		\caption{Band structure as a function of $\kappa_1$ in the SSH model with open boundary. Here, the total number of the waveguide is $N=30, \kappa_2=1, \kappa_{\delta} =0.5, \Delta=2$. (a) and (b) are the real and imaginary parts of eigenvalues. (c) and (d) are spectra with a defect at the left-most site. }
		\label{fig.3}
	\end{figure}
	
	The open boundary spectra are given in Fig.~\ref{fig.3}. Two non-zero edge modes occur when $\kappa_2-\Delta<\kappa_1<\Delta-\kappa_2$ in Fig.~\ref{fig.3}(a) and (b), suggesting the ordinary bulk-boundary correspondence is valid in our non-Hermitian anti-PT symmetric system. 
	As shown in Fig.~\ref{fig.3}(c), two new non-zero defect modes form (the red lines between $m_1$ and $m_2$, $m_1'$ and $m_2'$) when a defect is introduced in the left-most edge by replacing the coupling constant $\kappa_1$ with $\kappa_1+\kappa_{\delta}$ to test the robustness of the system. Clearly, the defect modes do not break the topological edge modes and the region of $\kappa_1$ is also unchanged~[Fig.~\ref{fig.3}(d)], proving that the edge modes are robust to the defect due to topological protection. 
	
  	To further investigate the energy distribution for the topological state, the stationary Schr\"{o}dinger-like equation
	\begin{gather}
		\label{eq:8}
		(E-\Delta)a_n=\kappa_1 b_n+\kappa_2 b_{n-1},\\
		\label{eq:9}
		(E+\Delta)b_n=-\kappa_1 a_n-\kappa_2 a_{n+1}.
	\end{gather}
	is used. As the band analysis predicts that the left edge mode has an eigenvalue of $E_{left}=\Delta$, with the boundary conditions
	\begin{gather}
		\label{eq:10}
		(E-\Delta)a_0=\kappa_1 b_0,\\
		\label{eq:11}
		(E+\Delta)b_0=-\kappa_1 a_0-\kappa_2 a_1.
	\end{gather}
	One can learn that
	\begin{gather}
		\label{eq:12}
		b_n=0,\\
		\label{eq:13}
		|a_n|^2=\left(\frac{\kappa_1}{\kappa_2}\right)^{2n}|a_0|^2.
	\end{gather}{\tiny }
   This indicates that the left incidence $a_0$ can induce a particular energy distribution $a_n$ and the energy only exists at the wide waveguide, site $A_n$. Peculiarly, a stable edge mode forms when $\kappa_1<\kappa_2$, which agrees with the topological edge state. When $\kappa_1=\kappa_2$, it is seen from Eq.~\ref{eq:13} $|a_n|^2=|a_0|^2$, indicating that the light energy distributes equally in all the wide wavegiudes $A_n$ for the case with undefined winding number [see Fig.~\ref{fig.2}(b2)]. Figuratively speaking, the topological state has been "copied" to all the waveguides $A_n$ for this special state. 
   
   	\begin{figure}%[h]
   	\centering
   	\fbox{\includegraphics[width=\linewidth]{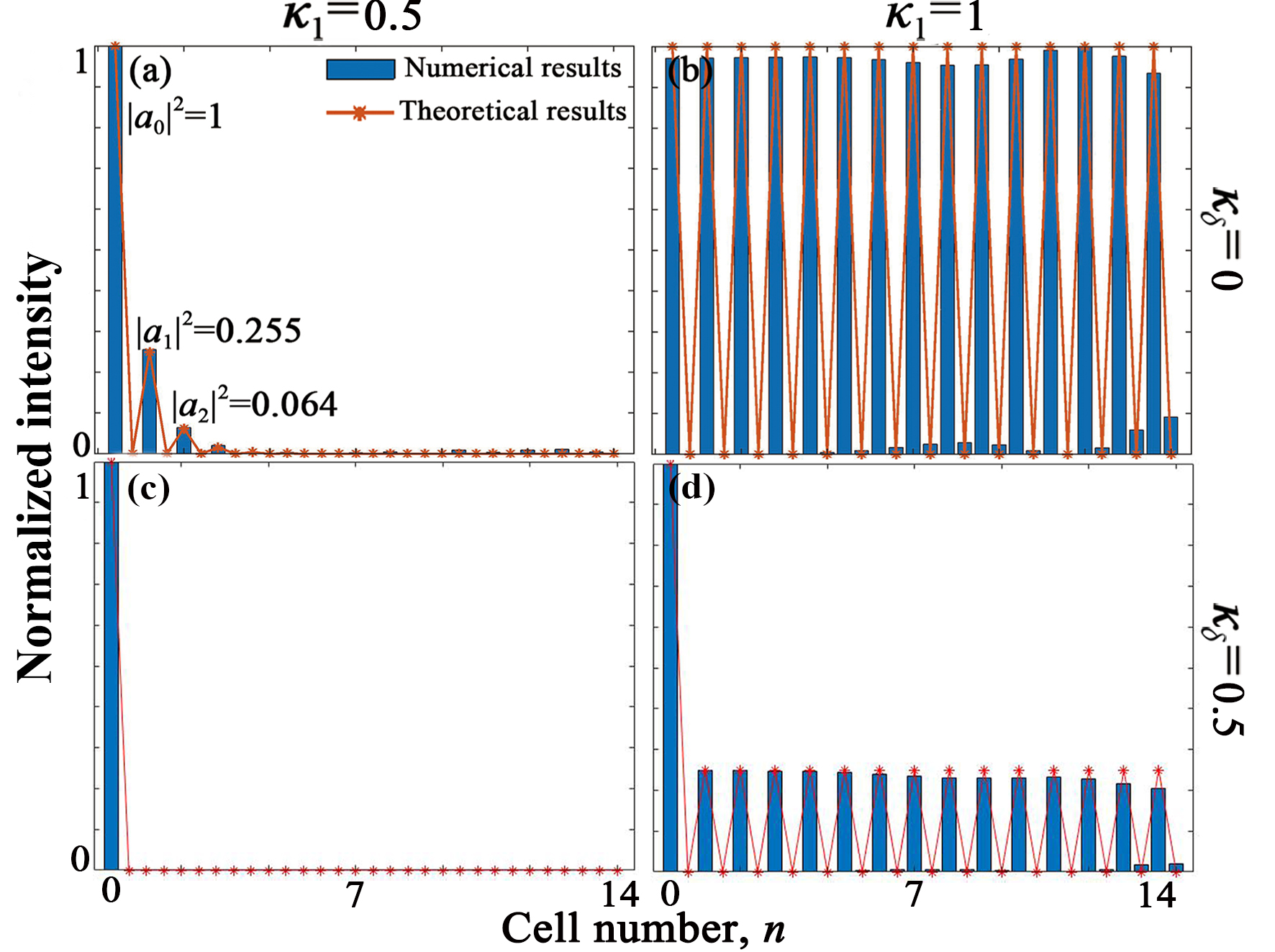}}
   	\caption{Numerical calculations of output energy distribution in both (a) topologically non-trivial($\kappa_1=0.5$) and (b) undefined cases($\kappa_1=1$). (c) and (d) add the defect at the edge with different $\kappa_1$ and same $\kappa_{\delta}$. }
   	\label{fig.4}
   \end{figure}
   
   Further considering the defect $\kappa_{\delta} \geq 0 $ in the first cell, one can get the following relations,
\begin{gather}
	\label{eq:14}
	(E-\Delta)a_0=(\kappa_1 + \kappa_{\delta}) b_0,\\
	\label{eq:15}
	(E+\Delta)b_0=-(\kappa_1 - \kappa_{\delta}) a_0-\kappa_2 a_1,
\end{gather}
   and obtain similar results with defect $\kappa_{\delta}$,
	\begin{gather}
	\label{eq:16}
	b_n=0,\\
	\label{eq:17}
	|a_n|^2=\left(\frac{\kappa_1}{\kappa_2}\right)^{2n}\left(\frac{\kappa_1-\kappa _{\delta}}{\kappa_1}\right)^{2}|a_0|^2.
\end{gather}{\tiny}
Considering  the modulated coupling constants should not break the lattice symmetry [see Fig.~\ref{fig.1}(c)], the coefficient term of $a_0$ on the right-hand side in Eq.\ref{eq:15} should satisfy $\kappa_1 - \kappa_{\delta} \geq 0$. Therefore the defect $0 \leq \kappa_{\delta} \leq \kappa_1$ is deduced. Since the imaginary part of the energy spectrum is zero for both the topological state [Fig.~\ref{fig.2}(a1) and (a2)] and undefined state  [Fig.~\ref{fig.2}(b1) and (b2)], it is easy to find from Eq.~\ref{eq:17} that $|a_n|^2\leq |a_0|^2$, indicating that the strongest light energy still distributes in the first waveguide $A_0$ and the light field in other waveguides $A_n$ can be controlled by adjusting the coupling constants $\kappa_1$, $\kappa_2$ and defect $\kappa_{\delta}$.

	To see the characteristics of the system more vividly, we calculated the output energy distribution both numerically and theoretically. As shown in Fig. \ref{fig.4}, the energy will only reside on the sites $A_n$ when the left edge mode is excited for $\kappa_1<\kappa_2$~[see Fig.~\ref{fig.4}(a)]. The edge state will be strengthened when the defect is introduced ($\kappa_{\delta} = 0.5$), namely, the system is topologically nontrivial~[see Fig.~\ref{fig.4} (c)]. Interesting, it is found that all the light intensity has the same magnitudes when $\kappa_1=\kappa_2$ and $\kappa_{\delta} =0$, for which the winding number of the system is undefined~[see the inset in Fig.~\ref{fig.2} (b2)]. When $\kappa_1=\kappa_2$ and $\kappa_{\delta} =0.5$, the equal intensity distribution keeps except for the first waveguide $A_0$, this means that this kind of undefined state is also robust and be immune to the defect~[Fig.~\ref{fig.4}(d)]. 
	
	\begin{figure}[h]
		\centering
		\fbox{\includegraphics[width=\linewidth]{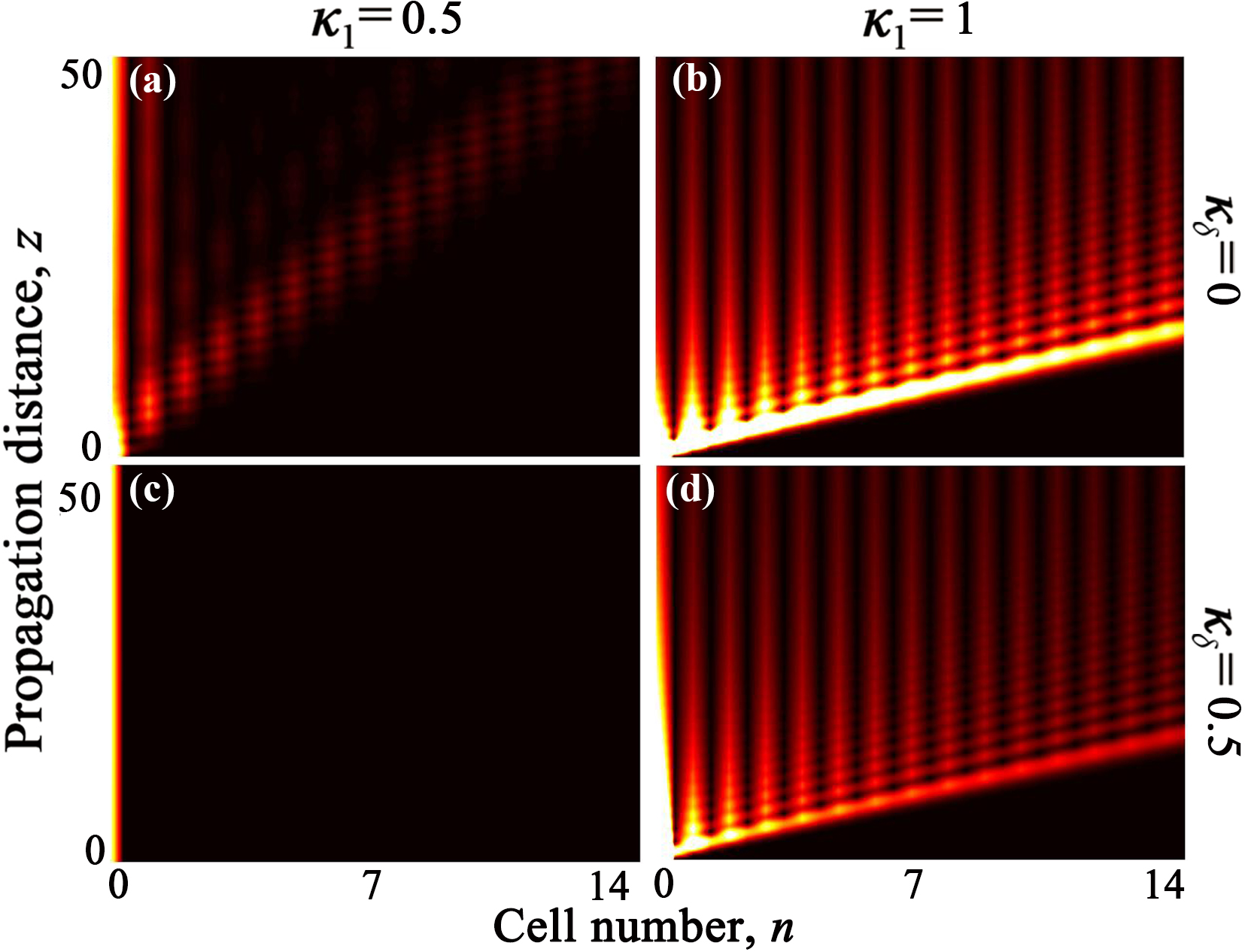}}
		\caption{Numerical simulation of light propagation incident from the left-most waveguide $A_0$.}
		\label{fig.5}
	\end{figure}
	
	Figure.~\ref{fig.5} shows the numerical simulation of light propagation in our system. Most of the light energy will be localized on waveguide $A_0$ and propagates along with it for an incident beam with a function of $ e^{-(\frac{n-n_0}{\omega_0})^2} $ ($\omega_0=0.1 $) when $\kappa_1<\kappa_2$ [see Fig.~\ref{fig.5}(a)]. When $\kappa_1=\kappa_2 =1$, the light will be coupled to the next nearest waveguides and finally propagates in all the sites $A_n$ ($n\in [0,14]$) with equal intensities [see Fig.~\ref{fig.5}(b)]. Contrastly, light wave is localized completely at $A_0$ when the defect is introduced ($\kappa_{\delta} = 0.5$) [see Fig.~\ref{fig.5}(c)]; This kind of strong localization could be attributed to the co-action of topological edge state and the defect mode as shown in Fig.~\ref{fig.3}(c) and Fig.~\ref{fig.4}(c). For the case of $\kappa_1 =1$ and $\kappa_{\delta} =0.5$ [Fig.~\ref{fig.5}(d)], the light will also distribute evenly in all the $A_n$ waveguides except that the first waveguide has larger energy due to the defect. Apparently, the light propagation is in good agreement with the former theoretical and numerical results.

	An applicable experimental proposal could be performed with the setup shown in Ref.~\cite{27koji2014,28Xia2021}. A femtosecond laser pulse @1064 nm is focused through a micro-objective lens with 20 times magnification. The wide(site $A_n$) and narrow(site $B_n$) waveguides could be induced in the fused silica glass by moving the optical platform relatively. The width and coupling coefficient of the waveguide can be modulated by controlling the moving velocity and the power of the femtosecond laser. A 632.8 nm He-Ne laser can be used as the excitation light source to probe the light dynamics.
	
	In summary, we present an anti-PT symmetric topological photonic lattice with perturbations. The topological edge state and undefined state can both be achieved in such an anti-PT photonic system by tuning the coupling constant. When the coupling constant $\kappa_1<\kappa_2$, most light transmits at the edge site $A_0$, forming a topological state. It is also found that the light distributes equally on all the wide waveguides $A_n$  when $\kappa_1=\kappa_2$, corresponding to a winding number undefined state. Further study shows the topological edge state will be immune to the defect and might be helpful to the formation of a more localized edge state, while the topologically undefined state is also robust to the defect except for the first waveguide where the input light is incident. Our work enriches the study of topologically undefined states and may offer important applications in optical circuits and optical communications in the future.
	
	\section{Funding Information}
This work is supported by the Open fund of Shandong Provincial Key Laboratory of Optics and Photonic Devices(K202008), Open Research Fund of State Key Laboratory of laser Interaction with Matter(SKLLIM1812), and Natural Science Foundation of Shaanxi Province, China(2017JM6014).

	\section{Acknowlege}
	The author Kaiwen Ji would like to acknowledge the support from China scholarship council (CSC No. 202006970015)

\section{Disclosures}
The authors declare no conflicts of interest.

	\bibliography{ref}

\begin{thebibliography}{10}
\newcommand{\enquote}[1]{``#1''}

\bibitem{01Markus2007}
M.~K{\"o}nig, S.~Wiedmann, C.~Br{\"u}ne, A.~Roth, H.~Buhmann, L.~W. Molenkamp,
  X.-L. Qi, and S.-C. Zhang, {\protect\JournalTitle{Science}} \textbf{318}, 766
  (2007).

\bibitem{02Petrifmmode2020}
J.~Petr\'a\ifmmode~\check{c}\else \v{c}\fi{}ek and V.~Kuzmiak,
  {\protect\JournalTitle{Phys. Rev. A}} \textbf{101}, 033805 (2020).

\bibitem{03ozawa_topological_2019}
T.~Ozawa, H.~M. Price, A.~Amo, N.~Goldman, M.~Hafezi, L.~Lu, M.~C. Rechtsman,
  D.~Schuster, J.~Simon, O.~Zilberberg, and I.~Carusotto,
  {\protect\JournalTitle{Reviews of Modern Physics}} \textbf{91}, 015006
  (2019).

\bibitem{04mukherjee_experimental_2017}
S.~Mukherjee, A.~Spracklen, M.~Valiente, E.~Andersson, P.~$\ddot{O}$hberg,
  N.~Goldman, and R.~R. Thomson, {\protect\JournalTitle{Nature Communications}}
  \textbf{8}, 13918 (2017).

\bibitem{05gao_probing_2016}
F.~Gao, Z.~Gao, X.~Shi, Z.~Yang, X.~Lin, H.~Xu, J.~D. Joannopoulos,
  M.~Solja\'ci\v{c}, H.~Chen, L.~Lu, Y.~Chong, and B.~Zhang,
  {\protect\JournalTitle{Nature Communications}} \textbf{7}, 11619 (2016).

\bibitem{06cheng_robust_2016}
X.~Cheng, C.~Jouvaud, X.~Ni, S.~H. Mousavi, A.~Z. Genack, and A.~B. Khanikaev,
  {\protect\JournalTitle{Nature Materials}} \textbf{15}, 542 (2016).

\bibitem{07wu_scheme_2015}
L.-H. Wu and X.~Hu, {\protect\JournalTitle{Physical Review Letters}}
  \textbf{114}, 223901 (2015).

\bibitem{08zhang2020}
L.~Zhang, Y.~Yang, Z.-K. Lin, P.~Qin, Q.~Chen, F.~Gao, E.~Li, J.-H. Jiang,
  B.~Zhang, and H.~Chen, {\protect\JournalTitle{Advanced Science}} \textbf{7},
  1902724 (2020).

\bibitem{09yu2021}
L.~Yu, H.~Xue, and B.~Zhang, {\protect\JournalTitle{Applied Physics Letters}}
  \textbf{118}, 071102 (2021).

\bibitem{10Zhou2017}
X.~Zhou, Y.~Wang, D.~Leykam, and Y.~D. Chong, {\protect\JournalTitle{New
  Journal of Physics}} \textbf{19}, 095002 (2017).

\bibitem{11Chern1993Hatsugai}
Y.~Hatsugai, {\protect\JournalTitle{Phys. Rev. Lett.}} \textbf{71}, 3697
  (1993).

\bibitem{12Anomalous2016Lee}
T.~E. Lee, {\protect\JournalTitle{Phys. Rev. Lett.}} \textbf{116}, 133903
  (2016).

\bibitem{13liangtopological2013}
S.-D. Liang and G.-Y. Huang, {\protect\JournalTitle{Physical Review A}}
  \textbf{87}, 012118 (2013).

\bibitem{14Edge2018Yao}
S.~Yao and Z.~Wang, {\protect\JournalTitle{Phys. Rev. Lett.}} \textbf{121},
  086803 (2018).

\bibitem{15Bulk2019Jin}
L.~Jin and Z.~Song, {\protect\JournalTitle{Phys. Rev. B}} \textbf{99}, 081103
  (2019).

\bibitem{16quandt_winding_2021}
A.~Quandt, {\protect\JournalTitle{Physica B: Condensed Matter}} \textbf{612},
  412867 (2021).

\bibitem{17Wang2020}
W.~Wang, W.~Gao, L.~Cao, Y.~Xiang, and S.~Zhang, {\protect\JournalTitle{Light:
  Science {\&} Applications}} \textbf{9}, 40 (2020).

\bibitem{18Yao2018}
S.~Yao, F.~Song, and Z.~Wang, {\protect\JournalTitle{Phys. Rev. Lett.}}
  \textbf{121}, 136802 (2018).

\bibitem{19Borgnia2020}
D.~S. Borgnia, A.~J. Kruchkov, and R.-J. Slager, {\protect\JournalTitle{Phys.
  Rev. Lett.}} \textbf{124}, 056802 (2020).

\bibitem{20Sone2020}
K.~Sone, Y.~Ashida, and T.~Sagawa, {\protect\JournalTitle{Nature
  Communications}} \textbf{11}, 5745 (2020).

\bibitem{21Anomalous2020Gao}
P.~Gao, M.~Willatzen, and J.~Christensen, {\protect\JournalTitle{Phys. Rev.
  Lett.}} \textbf{125}, 206402 (2020).

\bibitem{22Asboth2016}
J.~K. Asb\'{o}th, L.~Oroszl\'{a}ny, and A.~P\'{a}lyi, \emph{A {Short} {Course}
  on {Topological} {Insulators}}, vol. 919 of \emph{Lecture {Notes} in
  {Physics}} (Springer International Publishing, Cham).

\bibitem{23Lu2014}
L.~Lu, J.~D. Joannopoulos, and M.~Solja{\v{c}}i{\'{c}},
  {\protect\JournalTitle{Nature Photonics}} \textbf{8}, 821 (2014).

\bibitem{24Ji2018}
K.~Ji, Z.~Wen, Z.~Liu, Y.~Dai, K.~Han, P.~Gao, A.~Gao, J.~Bai, G.~Zhang, and
  X.~Qi, {\protect\JournalTitle{Opt. Lett.}} \textbf{43}, 4457 (2018).

\bibitem{25Ji2020}
K.~Ji, Z.~Liu, Y.~Dai, Z.~Wen, Y.~Wang, G.~Zhang, J.~Bai, and X.~Qi,
  {\protect\JournalTitle{Opt. Lett.}} \textbf{45}, 49 (2020).

\bibitem{26PhysRevB.95.014201}
S.~Longhi, {\protect\JournalTitle{Physical Review B}} \textbf{95}, 014201
  (2017).

\bibitem{28Xia2021}
S.~Xia, D.~Kaltsas, D.~Song, I.~Komis, J.~Xu, A.~Szameit, H.~Buljan, K.~G.
  Makris, and Z.~Chen, {\protect\JournalTitle{Science}} \textbf{372}, 72
  (2021).

\bibitem{27koji2014}
K.~Sugioka and Y.~Cheng, {\protect\JournalTitle{Light: Science {\&}
  Applications}} \textbf{3}, e149 (2014).

\end{thebibliography}
	
	% Full bibliography added automatically for Optics Letters submissions; the following line will simply be ignored if submitting to other journals.
	%Note that this extra page will not count against page length
	\bibliographyfullrefs{ref}
	
	%Manual citation listc
	%\begin{thebibliography}{1}
	%\bibitem{Zhang:14}
	%Y.~Zhang, S.~Qiao, L.~Sun, Q.~W. Shi, W.~Huang, %L.~Li, and Z.~Yang,
	 %\enquote{Photoinduced active terahertz metamaterials with nanostructured
	%anadium dioxide film deposited by sol-gel method,} Opt. Express \textbf{22},
	%11070--11078 (2014).
	%\end{thebibliography}
	
	% Please include bios and photos of all authors for aop articles
	\ifthenelse{\equal{\journalref}{aop}}{%
		\section*{Author Biographies}
		\begingroup
		\setlength\intextsep{0pt}
		\begin{minipage}[t][6.3cm][t]{1.0\textwidth} % Adjust height [6.3cm] as required for separation of bio photos.
			\begin{wrapfigure}{L}{0.25\textwidth}
				\includegraphics[width=0.25\textwidth]{john_smith.eps}
			\end{wrapfigure}
			\noindent
			{\bfseries John Smith} received his BSc (Mathematics) in 2000 from The University of Maryland. His research interests include lasers and optics.
		\end{minipage}
		\begin{minipage}{1.0\textwidth}
			\begin{wrapfigure}{L}{0.25\textwidth}
				\includegraphics[width=0.25\textwidth]{alice_smith.eps}
			\end{wrapfigure}
			\noindent
			{\bfseries Alice Smith} also received her BSc (Mathematics) in 2000 from The University of Maryland. Her research interests also include lasers and optics.
		\end{minipage}
		\endgroup
	}{}

\end{document}